\documentclass[aps,prl,citeautoscript,twocolumn,citeautoscript,showkeys,floatfix]{revtex4-1}
  \usepackage[utf8]{inputenc}
\usepackage[T1]{fontenc}
\usepackage{natbib}
\usepackage{url}
\usepackage{graphicx}
\usepackage{textcomp}
\usepackage{underscore}
\usepackage{float}
\usepackage{amsmath}
\usepackage[version=3]{mhchem}
\usepackage{hyperref}
\date{\today}
\title{manuscript}
\begin{document}

\title{Effects of Sublattice Symmetry and Frustration on Ionic Transport in Garnet Solid Electrolytes}

\author{Boris Kozinsky}
\email{Boris.Kozinsky@us.bosch.com}
\affiliation{Robert BOSCH LLC, Research and Technology Center North America, 255 Main St, Cambridge, Massachusetts 02142, United States}

\author{Sneha A. Akhade}
\affiliation{Robert BOSCH LLC, Research and Technology Center North America, 255 Main St, Cambridge, Massachusetts 02142, United States}

\author{Pierre Hirel}
\affiliation{Unit\'e Mat\'eriaux et Transformations, B\^at. C6, Universit\'e de Lille 1, 59655 Villeneuve d’Ascq, France}

\author{Adham Hashibon}
\affiliation{Fraunhofer Institute for Mechanics of Materials IWM, W\"ohlerstr. 11, 79108 Freiburg, Germany}

\author{Christian Elsässer}
\affiliation{Fraunhofer Institute for Mechanics of Materials IWM, W\"ohlerstr. 11, 79108 Freiburg, Germany}

\author{Prateek Mehta}
\affiliation{Robert BOSCH LLC, Research and Technology Center North America, 255 Main St, Cambridge, Massachusetts 02142, United States}

\author{Alan Logeat}
\affiliation{Robert Bosch GmbH, Corporate Sector Research and Advance Engineering, Robert-Bosch-Platz 1, 70839 Gerlingen-Schillerh\"ohe, Germany}

\author{Ulrich Eisele}
\affiliation{Robert Bosch GmbH, Corporate Sector Research and Advance Engineering, Robert-Bosch-Platz 1, 70839 Gerlingen-Schillerh\"ohe, Germany}

\date{\today}

\begin{abstract}
We use rigorous group-theoretic techniques and molecular dynamics to investigate the connection between structural symmetry and ionic conductivity in the garnet family of solid Li-ion electrolytes. We identify new ordered phases and order-disorder phase transitions that are relevant for conductivity optimization. Ionic transport in this materials family is controlled by the frustration of the Li sublattice caused by incommensurability with the host structure at non-integer Li concentrations, while ordered phases explain regions of sharply lower conductivity. Disorder is therefore predicted to be optimal for ionic transport in this and other conductor families with strong Li interaction.
\end{abstract}


\maketitle

Solid-state Li-ion batteries have generated keen interest as advanced energy storage systems with applications in portable electronics and electric vehicles \cite{scrosati-2010-lithium}. A highly conductive and stable solid ceramic ionic conductor can enable high-energy density by serving as a protection layer for Li-metal anodes, and a stable electrolyte for high-voltage cathodes. At the same time, replacing flammable organic electrolytes with an inorganic compound will remove the threat of thermal runaway catastrophes. In this context, solid-state inorganic Li conducting oxides with the garnet structure are currently considered to be among the top promising candidates for solid electrolytes due to a combination of their performance and stability properties. The garnet-like conductor \ce{Li5La3M2O12} (M = Nb, Ta) was first reported by Thangadurai \emph{et al.} \cite{thangadurai-2003-novel-fast} and to date an ionic conductivity as high as $\sigma = 3\times10^{-3} S/cm$ has been measured \cite{murugan-2007-fast-lithium} for the cubic \ce{Li7La3Zr2O12} garnet. The prototypical garnet structure is cubic (Ia-3d No. 230) with the chemical formula \ce{Li_$x$B3C2O12}. This structure is remarkably robust in regard to changes in cation composition (B=La, Ca, Ba, Sr, Y, Pr, Nd, Sm-Lu and C=Zr, Ta, Nb, Nd, Te, W) and is able to accommodate Li concentrations $x$ ranging from 3 to 7. There are two distinct crystallographic cation sites that Li can occupy: octahedral (oct) \emph{48g} and tetrahedral (tet) \emph{24d}. Other cations occupy the \emph{24c} (B) and \emph{16a} (C) sites, and oxygen occupies the \emph{96h} site. It has been recognized that Li site arrangements vary with Li content $x$, but there are conflicting reports of the trends coming from experimental measurements \cite{xie-2011-lithium-distr,ocallaghan-2008-switc-fast}. The complexity presents a difficulty in understanding both structure and ionic transport mechanisms. Unfortunately, X-ray diffraction can provide limited help in characterizing Li arrangements due to the weak scattering from Li, and one has to resort to complicated neutron scattering methods. The nature of ionic conduction is even less clear, as it is nearly impossible to observe directly, while at the same time a complex relationship between crystal structure and conductivity is evident. Multiple phases have been reported across the composition space with widely ranging conductivity values, which depend not only on stoichiometry but also preparation routes. For example, it was reported that \ce{Li7La3Zr2O12} is tetragonal (t-LLZ) with two orders of magnitude lower conductivity than the cubic (c-LLZ) polymorph \cite{awaka-2009-synth-li7la3}. Despite the promise and strong interest in this material there is limited understanding and consensus regarding the structure and transport mechanisms, particularly regarding the optimal Li vacancy concentration and site occupation \cite{thompson-2015-tale-two-sites}. 

In this work, we present a novel analysis method and use it to derive a comprehensive description of the atomic mechanisms governing garnet crystals, which is able to explain, in a predictive way, both structural and dynamic effects in the wide composition space. Our unifying description of this important materials class is enabled by a novel combination of computational methods that include systematic hierarchical space group symmetry analysis coupled with high-throughput atomistic computations and molecular dynamics. Structural optimization and temperature dependent transport was investigated with both Born-Oppenheimer ab-initio molecular dynamics (AIMD) using Quantum Espresso \cite{giannozzi-2009-quant-espres} and classical molecular dynamics (CMD) using LAMMPS \cite{plimpton-1995-fast-paral}. Details of the computational procedures are given in the Supplementary Information. We note that our analysis focuses only on single-crystal transport and neglects effects of grain boundaries which can be significant in sample measurements.

The arrangement of Li ions on the available crystallographic sites is arguably the main degree of freedom that controls both structure and transport characteristics of the garnet solid electrolytes. In order to identify the ground state Li orderings at each composition, we need to compute energies of all possible orderings. However, the complexity of even the primitive unit crystal cell makes it impossible:  in a simple cubic representation of LLZ there are 72 Li sites, of which 56 are occupied, resulting in $10^{20}$ possible arrangements. In order to deal with the enormous combinatorial space of configurations, we combine rigorous group-theoretic analysis for classifying Li site arrangements for each composition together with automated computations of total energies, in order to identify occupancy rules. The resulting complete mapping of lowest energy structures at each composition is then used to predict new ordered or disordered phases of garnets and to postulate the dependency of conductivity on composition.
The first step in our analysis is the determination that the occupancy ratio of Li among tet and oct sites is universal: it depends only on the Li concentration $x$ in the composition. To demonstrate this, for each value of $x$ we generate a set of $\sim$100 candidate configurations of various oct:tet occupancy ratios. Each structure is relaxed using DFT and the occupancy of the lowest energy configuration is automatically extracted using a topological partitioning algorithm. This procedure involves partitioning of the crystal by constructing a compact geometric network of non-overlapping polyhedra with O atoms as the vertices. Each polyhedron is then analyzed at each time step by a 3D convex hull algorithm to determine whether there is a Li ion present within the boundaries. The advantage of this topology-based approach is that it gives accurate indications of site occupancies, free from thermal noise, without the need of arbitrary distance-based metrics.

\begin{figure}[htb]
\centering
\includegraphics[width=3in]{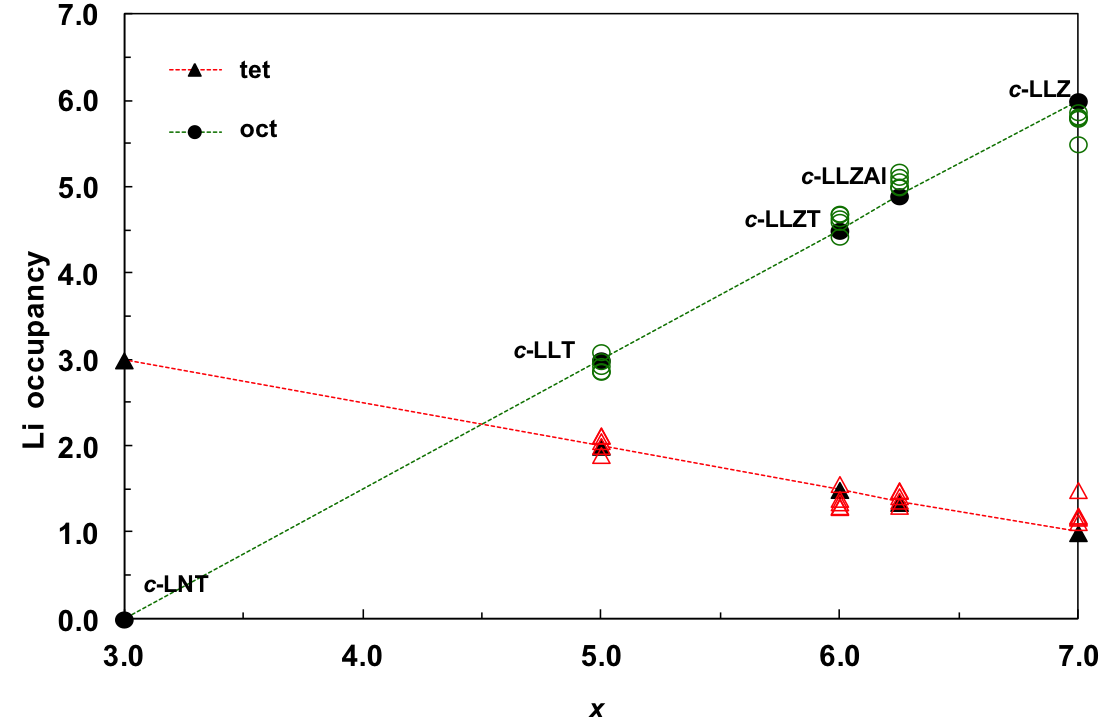}
\caption{Linear trend of Li occupancy (black) of oct and tet sites as a function of Li content. Temperature dependent oct (green) and tet (red) Li occupancy (500-1300 K) do not significantly deviate from occupancies at 0 K. Labels denote: c - cubic, LLZ - \ce{Li7La3Zr2O12}, LLZAl - \ce{Li_{6.25}Al_{0.75}La3Zr2O12}, LLZT - \ce{Li6La3ZrTaO12}, LLT - \ce{Li5La3Ta2O12}, LNT - \ce{Li3Nb3Te2O12} \label{fig:linear-trend}}
\end{figure}

The obtained occupancy trend is surprisingly linear, with Li favoring tet site occupancy when the Li content is low ($x < 4.5$). While qualitatively similar trends have been observed in several experimental reports \cite{xie-2011-lithium-distr,ocallaghan-2008-switc-fast,cussen-2010-struc,logeat-2012-from} quantitative results show a significant variation caused by experimental uncertainties and differences in material preparation. In order to investigate the effects of dynamics on this trend we performed AIMD at several temperatures. Fig.\ref{fig:linear-trend} shows the temperature dependent occupancy over different garnet compositions averaged over each AIMD run. It can be inferred that the trend is largely independent of temperature. The weak temperature dependence is in agreement with several in situ neutron diffraction experiments conducted on \ce{Li5La3Ta2O12} \cite{cussen-2006} and \ce{Li6BaLa2Ta2O12} \cite{ocallaghan-2007-lithium-li} between room temperature and up to 900 K that reported negligible variation in the lithium distribution across the tet and oct sites.

The occupancy constraint allows us to reduce the number of candidate configurations in the search for the ground state at each composition ($3 < x < 7$), but the number of possibilities is still on the order of $10^{19}$. To reduce the number of relevant configurations we perform a systematic search through crystallographic symmetry groups, subject to the universal constraint of the site occupancy ratio. This mathematical procedure enables us to efficiently analyze the entire space of possible ordered ground state structures in the garnet crystal family, using only a few hundred configurations. The group theoretic analysis procedure starts with crystallographic projections of the prototypical cubic symmetry group Ia$\bar{3}$d (no. 230) of the host lattice onto lower-symmetry subgroups in order to enumerate ordered structures at partial Li occupancies. The subgroup projection is implemented using tools provided by the Bilbao Crystallographic Server \cite{aroyo-2011-cryst}. We only consider cubic and tetragonal candidate subgroups, motivated by the experimental observation of only these two lattice types in the garnet family. First, for each Li concentration $x$, we determine all subgroups of group No.230 with the k-index up to 2. (The k-index indicates the multiplication factor relating the volume of the primitive cell of the subgroup with respect to the primitive cell of the original prototype structure.) This limit is imposed in order to limit the search to ordered structures with a reasonably compact description of a highly ordered phase. We then analyze the splittings \cite{kroumova-1998-wycks} of Wyckoff positions in each subgroup and conjugacy class in order to identify symmetrically related sets of Li sites and their multiplicities. The oct:tet occupancy ratio constraint is introduced by requiring that each resulting structure accommodates the correct ratio with the available Li site multiplicities. Finally, for each ordered structure that satisfies these constraints we perform a DFT variable-cell optimization calculations and identify the Li configuration of lowest energy for each composition $x$. The main result of this analysis is the non-trivial finding that for each of the integer values of $x=3,4,5,6,7$ there exist ordered configurations with cubic or tetragonal symmetry, and that some are more robust than others with respect to the tendency to disorder.
\begin{figure}[htb]
\centering
\includegraphics[width=3in]{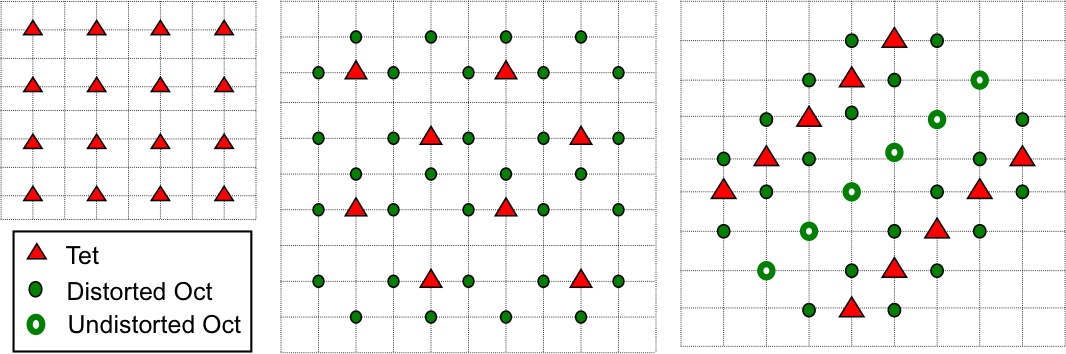}
\caption{2D representation of the ground-state Li arrangement for $x=3$ (left), $x=6$ (middle) and $x=7$ (right). Tet sites are denoted by grid intersections, and oct sites occupy the segment midpoints. The other constituents of the crystal are omitted for ease of viewing. \label{fig:2dmapping}}
\end{figure}

The topology of the Li site network determines the degree of ordering and consequently the transport characteristics in a non-trivial manner. Each tet site shares a face with four neighboring oct sites while each oct site shares a face with two neighboring tet sites. In order to easily visualize the Li arrangements, we map the topology of the 3D Li network onto a 2D grid (which provides an only locally correct mapping). Fig. \ref{fig:2dmapping} shows the 2D representations of the resulting ground-state configurations of Li concentrations $x$=3, 6 and 7. In the following we discuss in detail the Li configurations we identify for each integer concentration $x$. We note that it is not possible to find ordered structures (within the specified constraints) for off-integer concentrations, so they are predicted to be cubic and disordered.

\textbf{x=7.} From the subgroup projection procedure, the highest symmetry subgroup that allows for the occupancy ratio of 1:6 (tet:oct) is the tetragonal group I4$_{\text{1}}$/acd (No. 142). It is not possible to find any cubic subgroups of group No. 230, subject to the constraints of occupancy and unit cell size. DFT calculations identify the tetragonal layered structure as the lowest energy configuration. In this configuration Li occupies all the oct sites as well as 1/3 of the tet sites in a characteristic well-ordered layered arrangement (illustrated in Fig. \ref{fig:2dmapping} right), which imparts a noticeable tetragonal distortion (c/a=1.04 for t-LLZ) to the unit cell. XRD/Neutron diffraction results \cite{awaka-2009-synth-li7la3} confirm the tetragonal nature of the ground state structure of LLZ, as well as the layered Li atom arrangement. Li in an oct site (\emph{48g} in the prototype structure), is distorted to occupy the Wyckoff position \emph{96h} when one of the neighboring tet sites is occupied. This oct site distortion is present in other garnet compositions as well, indicating a significant degree of interaction among Li in the sublattice. The tetragonal structure is quite stationary, which is indicated by the high energy cost to move a Li ion from an oct site to a tet site ($\sim$0.7eV) \cite{meier-2014-solid-state-elect} and the tendency to rearrange subsequently back to the original state. In addition to the tetragonal structure, however, we find multiple low-symmetry configurations with the Li occupancy of 1.5:5.5 (tet:oct), which deviates slightly from the expected occupancy trend. These structures are higher in energy by as little as 5 meV per Li site and the cell shape is much closer to cubic, as expected from lower symmetry configurations. This is consistent with experimental observations that a phase transition can turn the ordered tetragonal LLZ (t-LLZ) to a disordered cubic form (c-LLZ) \cite{bernstein-2012-origin-struc}.

\textbf{x=6.} The occupancy trend predicts a tet:oct ratio 1.5:4.5 for this composition. A representative compound of this composition is \ce{Li6La3ZrMO12}, where M = \ce{Nb^5+}(LLZN) or \ce{Ta^5+}(LLZT). Symmetry subgroup analysis of Wyckoff position splittings identifies only one possible cubic group No. 198 (P2$_{1}$3). Within this space group the half-half mixing of cations (Zr:Ta / Zr:Nb) can also be accommodated in several ways. Overall there are 48 structures corresponding to arrangements of Li ions and cations within this space group, and we perform DFT variable-cell relaxation calculations to identify the ground state configuration for both mixed Zr-Ta and Zr-Nb cation compositions. The lowest energy structure exhibits a rock-salt type ordering of the 4+/5+ cations, and a simple pattern of Li ions, where each tet Li has exactly 3 oct neighbors (see Supporting Information). In addition, each tet Li has zero tet neighbors, and exactly half of the tetrahedral sites are occupied. (see Fig. \ref{fig:2dmapping} middle). The excitation energy to disturb the ordering is calculated to be 23 meV/Li site. Although experimental results report that $x=6$ compositions are indeed cubic \cite{awaka-2009-synth-li6cal-li6bal}, a detailed characterization of Li ion arrangement is presently lacking. Our finding presents new information explaining the likely ground state structure of this composition, which can anticipate confirmation from neutron diffraction studies of carefully annealed samples. 

\textbf{x=5.} The prescribed occupancy ratio of 2:3 (tet:oct) is possible to obtain within the tetragonal group No. 98 (I4$_1$22) as the highest symmetry compatible with the Li site Wyckoff position splitting. Group No. 80 (I4$_1$) is a tetragonal subgroup of No. 98, and also accommodates the required occupancy. DFT variable-cell relaxation calculations of \ce{Li5La3M2O12} (M = \ce{Nb^5+} or \ce{Ta^5+}) identify a tetragonal structure of group No. 98 as the lowest energy configuration. However, we also find multiple tetragonal structures corresponding to space group No. 80 that are only 9 meV per Li site higher in energy than the ground state. We predict that this degeneracy is likely to yield a statistically averaged cubic structure at room temperature with no long range Li order, as synthesized, for this composition \cite{peng-2013-low-li5la3}. We propose that it may be possible to detect tetragonal domains with sensitive neutron diffraction equipment, for samples annealed at lower temperatures. Similar to the $x=6$ case, we find that the relative energies of arrangements in\ce{Li5La3Ta2O12} (LLT) and \ce{Li5La3Nb2O12} (LLN) garnets are very close, predicting similar structural and transport behavior in these compositions.

\textbf{x=4.}  The tet:oct occupancy ratio of 2.5:1.5 can be accommodated by Wyckoff position splitting in tetragonal groups No. 117, 116, and 95. By computing total energies for each of 142 possible optimized configurations we find that the lowest energy structures belong to space group No. 95. The one identified as the lowest energy ordered structure is detailed in the Supporting Information. The next-lowest energy structure in this family is higher in energy by 20 meV/Li site, which also makes it challenging to observe especially if the high-temperature synthesis procedure involves a rapid quenching step. Unless carefully annealed, this composition is likely to be cubic and disordered, as synthesized.

\textbf{x=3.} In the classical silicate garnet structure (space group No. 230) all the \emph{24d} sites (tet) are occupied but none of the \emph{48g} (oct) sites. In the case of Li garnets, our subgroup projection approach identifies a unique structure with group No. 230 as the highest symmetry compatible with full tet occupancy (as dictated by the occupancy trend in Fig. \ref{fig:linear-trend}). This well-ordered cubic structure is strongly favored energetically (shown in Fig. \ref{fig:2dmapping} left), requiring more than 1.0 eV of energy to displace a Li ion from its tet position. Previous neutron diffraction studies \cite{ocallaghan-2006-struc-ionic} have determined the lithium distribution in \ce{Li3LaTe2O12} and \ce{Li3Nd3W2O12} and conform to our prediction.

\begin{figure}[htb]
\centering
\includegraphics[width=3in]{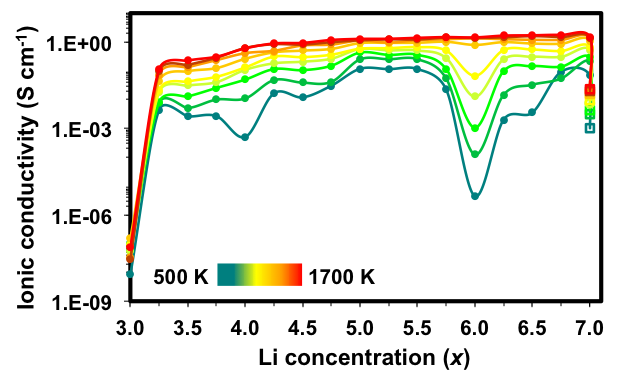}
\caption{Ionic conductivity (S/cm) of \ce{Li^{+}} for temperatures 500 K to 1700 K as a function of Li content $x$ in the garnet crystal, from CMD simulations. At $x=7$, the conductivity of the tetragonal (t-Li7) (denoted as hollow squares) and cubic (c-Li7) phases of \ce{Li7La3Zr2O12} have been depicted. \label{fig:composition}}
\end{figure}

The overall picture is that both tetragonal and cubic phases appear in the composition range, but the ionic conductivity depends primarily on the degree and robustness of order in the Li sublattice. Results on Fig. \ref{fig:composition} indicate that the conductivity, as a function of composition, exhibits a non-trivial trend where it is strongly suppressed by the presence of a stable ordered Li arrangement. In the structure analysis above we identified $x=3, 4, 6$ and $x=7$ (t-LLZ) as having well-ordered stable ground state configurations in the Li sub-lattice (Fig. \ref{fig:2dmapping}) with non-negligible excitation energies. Li conduction within an ordered sublattice involves creation of defects, which costs energy. On the other hand, much shallower energy landscapes results in better ionic mobility in disordered configurations that arise in situations of frustration. These are exactly the compositions, such as half-integral $x$, where it is mathematically impossible to find a simple ground state with a symmetry arrangement compatible with the host crystal lattice. As a result the Li network is intrinsically frustrated, disordered and mobile. We note that the interplay between possible symmetries of carrier configurations and conductivity is reminiscent to the order-disorder controlled transport observed in frustrated models, such as the triangular 2D anti-ferromagnetic Ising model \cite{novikov-2005-correl}.

\begin{figure}[htb]
\centering
\includegraphics[width=3in]{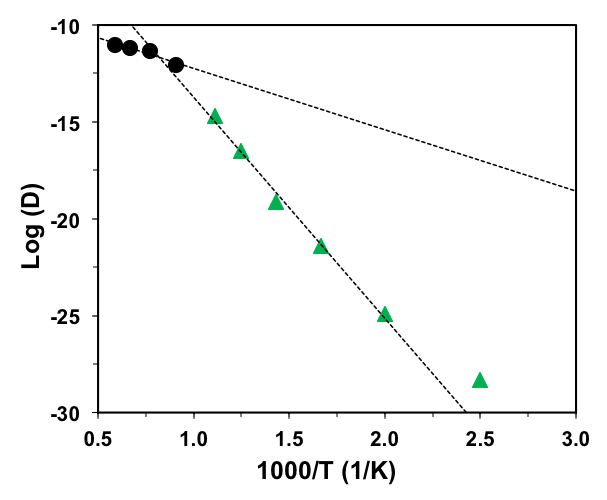}
\caption{Diffusion coefficient versus inverse temperature for the $x=6$ garnet structure, illustrating the order-disorder phase transition at $\sim$900K, as computed using CMD. Black circles represent high-temperature data and green triangles represent low-temperature data \label{fig:arrhenius}}
\end{figure}

In the case of $x=6$, the newly identified stable ordered ground state structure results in poor mobility at lower temperatures. As depicted in Fig. \ref{fig:composition}, a decline of ionic motion at even higher temperatures up to 900 K is indicative of Li sublattice configuration freezing, which is confirmed by topological structural analysis. The Arrhenius plot (Fig. \ref{fig:arrhenius}) of diffusion coefficients across the temperature range of the CMD simulations shows a clear indication of a sublattice phase transition where the Li-ion arrangement at low temperatures freezes into its ground-state configuration at lower temperatures, while at high temperature the sublattice melts and disorder prevails. A relatively low activation energy $E_A$ of 0.27 eV is obtained for high temperatures (1700-1100 K). But for a range of lower temperatures (900-500 K), the extracted $E_A$ is as high as 0.98 eV. We expect the actual experimental activation energy to be within these limits depending on the sample preparation procedure of the $x=6$ crystal, that may result in a partially locked Li arrangement. For concentrations just above and below $x=6$, mobile defects move freely in the background of the stable configuration, leading to sharp increases of conductivity. In this picture the concentration of free carriers remain smaller that the nominal Li concentration $x$ in this and other composition regions close to an ordered ground state. This picture explains the recent finding that only about 10\% of Li in garnets contribute to conductivity \cite{nozaki-2014-li-qens}, and also suggests that half-integer concentration are optimal for conductivity. For $x=3$ and $x=7$ (t-LLZ) the CMD simulations do not show any significant ionic motion even at high temperature. Indeed, for $x=3$, an ordered disconnected network of only tet Li cannot move because migration to an oct site is very unfavorable. As more Li ions are introduced, they act as free carriers moving in the background of a static high-symmetry $x=3$ configuration, leading to a rapid increase in conductivity. A similar situation prevails for the fully filled set of oct sites in the ordered tetragonal ground state of $x=7$ (t-LLZ), where the cost of creating oct vacancies is high. Reducing Li content by adjusting the composition introduces mobile vacancies that increase conductivity. At the same time, a higher degree of motion can also be observed in the ``excited'' configuration of LLZ with cubic structure c-LLZ and a tet:oct occupancy ratio of 1.5:5.5. In this case oct vacancies are present and act as carriers, and the availability of multiple degenerate configurations leads to disorder and significant freedom of motion. The simulations predict a distinct difference of 1-2 orders of magnitude in the conductivity between the ground state (t-Li7) and the disordered state (c-Li7), consistent with experimental observations \cite{awaka-2011-cryst-struc}. CMD simulations for the case of $x=5$ exhibit better transport, as compared to the other integer compositions, due to a greater intrinsic propensity towards disorder. This can be attributed to the presence of multiple configurations that are energetically similar to the ground-state arrangement and tend to yield an average low-symmetry configuration. The slight dip of conductivity at $x=4$ is again consistent with the presence of an ordered ground state; however, in this case disorder is seen to set in at lower temperatures than for $x=6$.

In conclusion, we provide the basis for understanding the non-trivial relationship between composition, structure and transport that governs performance of garnets in the wide range of compositions. Through novel systematic structural symmetry analysis based on the universal occupancy trend, combined with ab-initio energy computations, we predict that ordered ground state crystal structures exist for integer compositions. Some of these structures have been confirmed experimentally, while others are yet to be detected. In particular, we identify ordered ground states at most integer compositions (notably new phases for the $x=4$ and $x=6$), and characterize their configurational excitations. The lack of stable ordered states at half-integer concentrations results in disordered frustrated configurations that maximize carrier concentration and mobility. This is a consequence of the strong Li-Li interaction and the particular symmetry structure of the host lattice, and indicates that it is primarily the Li concentration and arrangement that controls transport in garnets. Based on this new understanding, we expect that intrinsic conductivity is highest at compositions around $x=5.5$ and $x=6.5$, and that is where future experimental efforts should focus. We speculate that presented methods and understanding are relevant to other families of ionic conductors, where Li concentration is adjustable and Li-Li interaction is strong, such that sublattice frustration and disorder can be used to optimize ionic transport.

\begin{acknowledgments}
This research used resources of the Oak Ridge Leadership Computing Facility located in the Oak Ridge National Laboratory, which is supported by the Office of Science of the Department of Energy under Contract DE-AC05-00OR22725. The part of this work done at Robert Bosch GmbH and Fraunhofer IWM in Germany was funded by the German Federal Ministry of Education and Research BMBF within the Project HE-Lion (Grant No. INLB03088008). 
\end{acknowledgments}

\bibliography{references}

\begin{thebibliography}{23}%
\makeatletter
\providecommand \@ifxundefined [1]{%
 \@ifx{#1\undefined}
}%
\providecommand \@ifnum [1]{%
 \ifnum #1\expandafter \@firstoftwo
 \else \expandafter \@secondoftwo
 \fi
}%
\providecommand \@ifx [1]{%
 \ifx #1\expandafter \@firstoftwo
 \else \expandafter \@secondoftwo
 \fi
}%
\providecommand \natexlab [1]{#1}%
\providecommand \enquote  [1]{``#1''}%
\providecommand \bibnamefont  [1]{#1}%
\providecommand \bibfnamefont [1]{#1}%
\providecommand \citenamefont [1]{#1}%
\providecommand \href@noop [0]{\@secondoftwo}%
\providecommand \href [0]{\begingroup \@sanitize@url \@href}%
\providecommand \@href[1]{\@@startlink{#1}\@@href}%
\providecommand \@@href[1]{\endgroup#1\@@endlink}%
\providecommand \@sanitize@url [0]{\catcode `\\12\catcode `\$12\catcode
  `\&12\catcode `\#12\catcode `\^12\catcode `\_12\catcode `\%12\relax}%
\providecommand \@@startlink[1]{}%
\providecommand \@@endlink[0]{}%
\providecommand \url  [0]{\begingroup\@sanitize@url \@url }%
\providecommand \@url [1]{\endgroup\@href {#1}{\urlprefix }}%
\providecommand \urlprefix  [0]{URL }%
\providecommand \Eprint [0]{\href }%
\providecommand \doibase [0]{http://dx.doi.org/}%
\providecommand \selectlanguage [0]{\@gobble}%
\providecommand \bibinfo  [0]{\@secondoftwo}%
\providecommand \bibfield  [0]{\@secondoftwo}%
\providecommand \translation [1]{[#1]}%
\providecommand \BibitemOpen [0]{}%
\providecommand \bibitemStop [0]{}%
\providecommand \bibitemNoStop [0]{.\EOS\space}%
\providecommand \EOS [0]{\spacefactor3000\relax}%
\providecommand \BibitemShut  [1]{\csname bibitem#1\endcsname}%
\let\auto@bib@innerbib\@empty
\bibitem [{\citenamefont {Scrosati}\ and\ \citenamefont
  {Garche}(2010)}]{scrosati-2010-lithium}%
  \BibitemOpen
  \bibfield  {author} {\bibinfo {author} {\bibfnamefont {B.}~\bibnamefont
  {Scrosati}}\ and\ \bibinfo {author} {\bibfnamefont {J.}~\bibnamefont
  {Garche}},\ }\href {\doibase 10.1016/j.jpowsour.2009.11.048} {\bibfield
  {journal} {\bibinfo  {journal} {Journal of Power Sources}\ }\textbf {\bibinfo
  {volume} {195}},\ \bibinfo {pages} {2419} (\bibinfo {year}
  {2010})}\BibitemShut {NoStop}%
\bibitem [{\citenamefont {Thangadurai}\ \emph {et~al.}(2003)\citenamefont
  {Thangadurai}, \citenamefont {Kaack},\ and\ \citenamefont
  {Weppner}}]{thangadurai-2003-novel-fast}%
  \BibitemOpen
  \bibfield  {author} {\bibinfo {author} {\bibfnamefont {V.}~\bibnamefont
  {Thangadurai}}, \bibinfo {author} {\bibfnamefont {H.}~\bibnamefont {Kaack}},
  \ and\ \bibinfo {author} {\bibfnamefont {W.~J.~F.}\ \bibnamefont {Weppner}},\
  }\href {\doibase 10.1111/j.1151-2916.2003.tb03318.x} {\bibfield  {journal}
  {\bibinfo  {journal} {Journal of the American Ceramic Society}\ }\textbf
  {\bibinfo {volume} {86}},\ \bibinfo {pages} {437} (\bibinfo {year}
  {2003})}\BibitemShut {NoStop}%
\bibitem [{\citenamefont {Murugan}\ \emph {et~al.}(2007)\citenamefont
  {Murugan}, \citenamefont {Thangadurai},\ and\ \citenamefont
  {Weppner}}]{murugan-2007-fast-lithium}%
  \BibitemOpen
  \bibfield  {author} {\bibinfo {author} {\bibfnamefont {R.}~\bibnamefont
  {Murugan}}, \bibinfo {author} {\bibfnamefont {V.}~\bibnamefont
  {Thangadurai}}, \ and\ \bibinfo {author} {\bibfnamefont {W.}~\bibnamefont
  {Weppner}},\ }\href {\doibase 10.1002/anie.200701144} {\bibfield  {journal}
  {\bibinfo  {journal} {Angewandte Chemie International Edition}\ }\textbf
  {\bibinfo {volume} {46}},\ \bibinfo {pages} {7778} (\bibinfo {year}
  {2007})}\BibitemShut {NoStop}%
\bibitem [{\citenamefont {Xie}\ \emph {et~al.}(2011)\citenamefont {Xie},
  \citenamefont {Alonso}, \citenamefont {Li}, \citenamefont
  {Fern{\'a}ndez-D{\'i}az},\ and\ \citenamefont
  {Goodenough}}]{xie-2011-lithium-distr}%
  \BibitemOpen
  \bibfield  {author} {\bibinfo {author} {\bibfnamefont {H.}~\bibnamefont
  {Xie}}, \bibinfo {author} {\bibfnamefont {J.~A.}\ \bibnamefont {Alonso}},
  \bibinfo {author} {\bibfnamefont {Y.}~\bibnamefont {Li}}, \bibinfo {author}
  {\bibfnamefont {M.~T.}\ \bibnamefont {Fern{\'a}ndez-D{\'i}az}}, \ and\
  \bibinfo {author} {\bibfnamefont {J.~B.}\ \bibnamefont {Goodenough}},\ }\href
  {\doibase 10.1021/cm201671k} {\bibfield  {journal} {\bibinfo  {journal}
  {Chemistry of Materials}\ }\textbf {\bibinfo {volume} {23}},\ \bibinfo
  {pages} {3587} (\bibinfo {year} {2011})}\BibitemShut {NoStop}%
\bibitem [{\citenamefont {O'Callaghan}\ \emph {et~al.}(2008)\citenamefont
  {O'Callaghan}, \citenamefont {Powell}, \citenamefont {Titman}, \citenamefont
  {Chen},\ and\ \citenamefont {Cussen}}]{ocallaghan-2008-switc-fast}%
  \BibitemOpen
  \bibfield  {author} {\bibinfo {author} {\bibfnamefont {M.~P.}\ \bibnamefont
  {O'Callaghan}}, \bibinfo {author} {\bibfnamefont {A.~S.}\ \bibnamefont
  {Powell}}, \bibinfo {author} {\bibfnamefont {J.~J.}\ \bibnamefont {Titman}},
  \bibinfo {author} {\bibfnamefont {G.~Z.}\ \bibnamefont {Chen}}, \ and\
  \bibinfo {author} {\bibfnamefont {E.~J.}\ \bibnamefont {Cussen}},\ }\href
  {\doibase 10.1021/cm703677q} {\bibfield  {journal} {\bibinfo  {journal}
  {Chemistry of Materials}\ }\textbf {\bibinfo {volume} {20}},\ \bibinfo
  {pages} {2360} (\bibinfo {year} {2008})}\BibitemShut {NoStop}%
\bibitem [{\citenamefont {Awaka}\ \emph
  {et~al.}(2009{\natexlab{a}})\citenamefont {Awaka}, \citenamefont {Kijima},
  \citenamefont {Hayakawa},\ and\ \citenamefont
  {Akimoto}}]{awaka-2009-synth-li7la3}%
  \BibitemOpen
  \bibfield  {author} {\bibinfo {author} {\bibfnamefont {J.}~\bibnamefont
  {Awaka}}, \bibinfo {author} {\bibfnamefont {N.}~\bibnamefont {Kijima}},
  \bibinfo {author} {\bibfnamefont {H.}~\bibnamefont {Hayakawa}}, \ and\
  \bibinfo {author} {\bibfnamefont {J.}~\bibnamefont {Akimoto}},\ }\href
  {\doibase 10.1016/j.jssc.2009.05.020} {\bibfield  {journal} {\bibinfo
  {journal} {Journal of Solid State Chemistry}\ }\textbf {\bibinfo {volume}
  {182}},\ \bibinfo {pages} {2046} (\bibinfo {year}
  {2009}{\natexlab{a}})}\BibitemShut {NoStop}%
\bibitem [{\citenamefont {Thompson}\ \emph {et~al.}(2015)\citenamefont
  {Thompson}, \citenamefont {Sharafi}, \citenamefont {Johannes}, \citenamefont
  {Huq}, \citenamefont {Allen}, \citenamefont {Wolfenstine},\ and\
  \citenamefont {Sakamoto}}]{thompson-2015-tale-two-sites}%
  \BibitemOpen
  \bibfield  {author} {\bibinfo {author} {\bibfnamefont {T.}~\bibnamefont
  {Thompson}}, \bibinfo {author} {\bibfnamefont {A.}~\bibnamefont {Sharafi}},
  \bibinfo {author} {\bibfnamefont {M.~D.}\ \bibnamefont {Johannes}}, \bibinfo
  {author} {\bibfnamefont {A.}~\bibnamefont {Huq}}, \bibinfo {author}
  {\bibfnamefont {J.~L.}\ \bibnamefont {Allen}}, \bibinfo {author}
  {\bibfnamefont {J.}~\bibnamefont {Wolfenstine}}, \ and\ \bibinfo {author}
  {\bibfnamefont {J.}~\bibnamefont {Sakamoto}},\ }\href {\doibase
  10.1002/aenm.201500096} {\bibfield  {journal} {\bibinfo  {journal} {Advanced
  Energy Materials}\ }\textbf {\bibinfo {volume} {5}},\ \bibinfo {pages}
  {1500096} (\bibinfo {year} {2015})}\BibitemShut {NoStop}%
\bibitem [{\citenamefont {Giannozzi}\ \emph {et~al.}(2009)\citenamefont
  {Giannozzi}, \citenamefont {Baroni}, \citenamefont {Bonini}, \citenamefont
  {Calandra}, \citenamefont {Car}, \citenamefont {Cavazzoni}, \citenamefont
  {Ceresoli}, \citenamefont {Chiarotti}, \citenamefont {Cococcioni},
  \citenamefont {Dabo}, \citenamefont {Corso}, \citenamefont {Gironcoli},
  \citenamefont {Fabris}, \citenamefont {Fratesi}, \citenamefont {Gebauer},
  \citenamefont {Gerstmann}, \citenamefont {Gougoussis}, \citenamefont
  {Kokalj}, \citenamefont {Lazzeri}, \citenamefont {Martin-Samos},
  \citenamefont {Marzari}, \citenamefont {Mauri}, \citenamefont {Mazzarello},
  \citenamefont {Paolini}, \citenamefont {Pasquarello}, \citenamefont
  {Paulatto}, \citenamefont {Sbraccia}, \citenamefont {Scandolo}, \citenamefont
  {Sclauzero}, \citenamefont {Seitsonen}, \citenamefont {Smogunov},
  \citenamefont {Umari},\ and\ \citenamefont
  {Wentzcovitch}}]{giannozzi-2009-quant-espres}%
  \BibitemOpen
  \bibfield  {author} {\bibinfo {author} {\bibfnamefont {P.}~\bibnamefont
  {Giannozzi}}, \bibinfo {author} {\bibfnamefont {S.}~\bibnamefont {Baroni}},
  \bibinfo {author} {\bibfnamefont {N.}~\bibnamefont {Bonini}}, \bibinfo
  {author} {\bibfnamefont {M.}~\bibnamefont {Calandra}}, \bibinfo {author}
  {\bibfnamefont {R.}~\bibnamefont {Car}}, \bibinfo {author} {\bibfnamefont
  {C.}~\bibnamefont {Cavazzoni}}, \bibinfo {author} {\bibfnamefont
  {D.}~\bibnamefont {Ceresoli}}, \bibinfo {author} {\bibfnamefont
  {L.}~\bibnamefont {Chiarotti}, \bibfnamefont {Guido}}, \bibinfo {author}
  {\bibfnamefont {M.}~\bibnamefont {Cococcioni}}, \bibinfo {author}
  {\bibfnamefont {I.}~\bibnamefont {Dabo}}, \bibinfo {author} {\bibfnamefont
  {A.~D.}\ \bibnamefont {Corso}}, \bibinfo {author} {\bibfnamefont {S.~d.}\
  \bibnamefont {Gironcoli}}, \bibinfo {author} {\bibfnamefont {S.}~\bibnamefont
  {Fabris}}, \bibinfo {author} {\bibfnamefont {G.}~\bibnamefont {Fratesi}},
  \bibinfo {author} {\bibfnamefont {R.}~\bibnamefont {Gebauer}}, \bibinfo
  {author} {\bibfnamefont {U.}~\bibnamefont {Gerstmann}}, \bibinfo {author}
  {\bibfnamefont {C.}~\bibnamefont {Gougoussis}}, \bibinfo {author}
  {\bibfnamefont {A.}~\bibnamefont {Kokalj}}, \bibinfo {author} {\bibfnamefont
  {M.}~\bibnamefont {Lazzeri}}, \bibinfo {author} {\bibfnamefont
  {L.}~\bibnamefont {Martin-Samos}}, \bibinfo {author} {\bibfnamefont
  {N.}~\bibnamefont {Marzari}}, \bibinfo {author} {\bibfnamefont
  {F.}~\bibnamefont {Mauri}}, \bibinfo {author} {\bibfnamefont
  {R.}~\bibnamefont {Mazzarello}}, \bibinfo {author} {\bibfnamefont
  {S.}~\bibnamefont {Paolini}}, \bibinfo {author} {\bibfnamefont
  {A.}~\bibnamefont {Pasquarello}}, \bibinfo {author} {\bibfnamefont
  {L.}~\bibnamefont {Paulatto}}, \bibinfo {author} {\bibfnamefont
  {C.}~\bibnamefont {Sbraccia}}, \bibinfo {author} {\bibfnamefont
  {S.}~\bibnamefont {Scandolo}}, \bibinfo {author} {\bibfnamefont
  {G.}~\bibnamefont {Sclauzero}}, \bibinfo {author} {\bibfnamefont {A.~P.}\
  \bibnamefont {Seitsonen}}, \bibinfo {author} {\bibfnamefont {A.}~\bibnamefont
  {Smogunov}}, \bibinfo {author} {\bibfnamefont {P.}~\bibnamefont {Umari}}, \
  and\ \bibinfo {author} {\bibfnamefont {R.~M.}\ \bibnamefont {Wentzcovitch}},\
  }\href {http://stacks.iop.org/0953-8984/21/i=39/a=395502} {\bibfield
  {journal} {\bibinfo  {journal} {Journal of Physics: Condensed Matter}\
  }\textbf {\bibinfo {volume} {21}},\ \bibinfo {pages} {395502} (\bibinfo
  {year} {2009})}\BibitemShut {NoStop}%
\bibitem [{\citenamefont {Plimpton}(1995)}]{plimpton-1995-fast-paral}%
  \BibitemOpen
  \bibfield  {author} {\bibinfo {author} {\bibfnamefont {S.}~\bibnamefont
  {Plimpton}},\ }\href {\doibase 10.1006/jcph.1995.1039} {\bibfield  {journal}
  {\bibinfo  {journal} {Journal of Computational Physics}\ }\textbf {\bibinfo
  {volume} {117}},\ \bibinfo {pages} {1} (\bibinfo {year} {1995})}\BibitemShut
  {NoStop}%
\bibitem [{\citenamefont {Cussen}(2010)}]{cussen-2010-struc}%
  \BibitemOpen
  \bibfield  {author} {\bibinfo {author} {\bibfnamefont {E.~J.}\ \bibnamefont
  {Cussen}},\ }\href {http://dx.doi.org/10.1039/B925553B} {\bibfield  {journal}
  {\bibinfo  {journal} {Journal of Materials Chemistry}\ }\textbf {\bibinfo
  {volume} {20}},\ \bibinfo {pages} {5167} (\bibinfo {year}
  {2010})}\BibitemShut {NoStop}%
\bibitem [{\citenamefont {Log\'eat}\ \emph {et~al.}(2012)\citenamefont
  {Log\'eat}, \citenamefont {K\"oohler}, \citenamefont {Eisele}, \citenamefont
  {Stiaszny}, \citenamefont {Harzer}, \citenamefont {Tovar}, \citenamefont
  {Senyshyn}, \citenamefont {Ehrenberg},\ and\ \citenamefont
  {Kozinsky}}]{logeat-2012-from}%
  \BibitemOpen
  \bibfield  {author} {\bibinfo {author} {\bibfnamefont {A.}~\bibnamefont
  {Log\'eat}}, \bibinfo {author} {\bibfnamefont {T.}~\bibnamefont {K\"oohler}},
  \bibinfo {author} {\bibfnamefont {U.}~\bibnamefont {Eisele}}, \bibinfo
  {author} {\bibfnamefont {B.}~\bibnamefont {Stiaszny}}, \bibinfo {author}
  {\bibfnamefont {A.}~\bibnamefont {Harzer}}, \bibinfo {author} {\bibfnamefont
  {M.}~\bibnamefont {Tovar}}, \bibinfo {author} {\bibfnamefont
  {A.}~\bibnamefont {Senyshyn}}, \bibinfo {author} {\bibfnamefont
  {H.}~\bibnamefont {Ehrenberg}}, \ and\ \bibinfo {author} {\bibfnamefont
  {B.}~\bibnamefont {Kozinsky}},\ }\href {\doibase 10.1016/j.ssi.2011.10.023}
  {\bibfield  {journal} {\bibinfo  {journal} {Solid State Ionics}\ }\textbf
  {\bibinfo {volume} {206}},\ \bibinfo {pages} {33} (\bibinfo {year}
  {2012})}\BibitemShut {NoStop}%
\bibitem [{\citenamefont {Cussen}(2006)}]{cussen-2006}%
  \BibitemOpen
  \bibfield  {author} {\bibinfo {author} {\bibfnamefont {E.~J.}\ \bibnamefont
  {Cussen}},\ }\href {\doibase 10.1039/b514640b} {\bibfield  {journal}
  {\bibinfo  {journal} {Chem. Commun.}\ }\textbf {\bibinfo {volume} {nil}},\
  \bibinfo {pages} {412} (\bibinfo {year} {2006})}\BibitemShut {NoStop}%
\bibitem [{\citenamefont {O'Callaghan}\ and\ \citenamefont
  {Cussen}(2007)}]{ocallaghan-2007-lithium-li}%
  \BibitemOpen
  \bibfield  {author} {\bibinfo {author} {\bibfnamefont {M.~P.}\ \bibnamefont
  {O'Callaghan}}\ and\ \bibinfo {author} {\bibfnamefont {E.~J.}\ \bibnamefont
  {Cussen}},\ }\href {http://dx.doi.org/10.1039/B700369B} {\bibfield  {journal}
  {\bibinfo  {journal} {Chemical Communications}\ }\textbf {\bibinfo {volume}
  {0}},\ \bibinfo {pages} {2048} (\bibinfo {year} {2007})}\BibitemShut
  {NoStop}%
\bibitem [{\citenamefont {Aroyo}\ \emph {et~al.}(2011)\citenamefont {Aroyo},
  \citenamefont {Perez-Mato}, \citenamefont {Orobengoa}, \citenamefont {Tasci},
  \citenamefont {de~la Flor},\ and\ \citenamefont {Kirov}}]{aroyo-2011-cryst}%
  \BibitemOpen
  \bibfield  {author} {\bibinfo {author} {\bibfnamefont {M.}~\bibnamefont
  {Aroyo}}, \bibinfo {author} {\bibfnamefont {J.}~\bibnamefont {Perez-Mato}},
  \bibinfo {author} {\bibfnamefont {D.}~\bibnamefont {Orobengoa}}, \bibinfo
  {author} {\bibfnamefont {E.}~\bibnamefont {Tasci}}, \bibinfo {author}
  {\bibfnamefont {G.}~\bibnamefont {de~la Flor}}, \ and\ \bibinfo {author}
  {\bibfnamefont {A.}~\bibnamefont {Kirov}},\ }\href@noop {} {\bibfield
  {journal} {\bibinfo  {journal} {Bulg Chem Commun}\ }\textbf {\bibinfo
  {volume} {43}},\ \bibinfo {pages} {183} (\bibinfo {year} {2011})}\BibitemShut
  {NoStop}%
\bibitem [{\citenamefont {Kroumova}\ \emph {et~al.}(1998)\citenamefont
  {Kroumova}, \citenamefont {Perez-Mato},\ and\ \citenamefont
  {Aroyo}}]{kroumova-1998-wycks}%
  \BibitemOpen
  \bibfield  {author} {\bibinfo {author} {\bibfnamefont {E.}~\bibnamefont
  {Kroumova}}, \bibinfo {author} {\bibfnamefont {J.~M.}\ \bibnamefont
  {Perez-Mato}}, \ and\ \bibinfo {author} {\bibfnamefont {M.~I.}\ \bibnamefont
  {Aroyo}},\ }\href {\doibase doi:10.1107/S0021889898005524} {\bibfield
  {journal} {\bibinfo  {journal} {Journal of Applied Crystallography}\ }\textbf
  {\bibinfo {volume} {31}},\ \bibinfo {pages} {646} (\bibinfo {year}
  {1998})}\BibitemShut {NoStop}%
\bibitem [{\citenamefont {Meier}\ \emph {et~al.}(2014)\citenamefont {Meier},
  \citenamefont {Laino},\ and\ \citenamefont
  {Curioni}}]{meier-2014-solid-state-elect}%
  \BibitemOpen
  \bibfield  {author} {\bibinfo {author} {\bibfnamefont {K.}~\bibnamefont
  {Meier}}, \bibinfo {author} {\bibfnamefont {T.}~\bibnamefont {Laino}}, \ and\
  \bibinfo {author} {\bibfnamefont {A.}~\bibnamefont {Curioni}},\ }\href
  {\doibase 10.1021/jp5002463} {\bibfield  {journal} {\bibinfo  {journal} {The
  Journal of Physical Chemistry C}\ }\textbf {\bibinfo {volume} {118}},\
  \bibinfo {pages} {6668} (\bibinfo {year} {2014})}\BibitemShut {NoStop}%
\bibitem [{\citenamefont {Bernstein}\ \emph {et~al.}(2012)\citenamefont
  {Bernstein}, \citenamefont {Johannes},\ and\ \citenamefont
  {Hoang}}]{bernstein-2012-origin-struc}%
  \BibitemOpen
  \bibfield  {author} {\bibinfo {author} {\bibfnamefont {N.}~\bibnamefont
  {Bernstein}}, \bibinfo {author} {\bibfnamefont {M.~D.}\ \bibnamefont
  {Johannes}}, \ and\ \bibinfo {author} {\bibfnamefont {K.}~\bibnamefont
  {Hoang}},\ }\href {http://link.aps.org/doi/10.1103/PhysRevLett.109.205702}
  {\bibfield  {journal} {\bibinfo  {journal} {Physical Review Letters}\
  }\textbf {\bibinfo {volume} {109}},\ \bibinfo {pages} {205702} (\bibinfo
  {year} {2012})}\BibitemShut {NoStop}%
\bibitem [{\citenamefont {Awaka}\ \emph
  {et~al.}(2009{\natexlab{b}})\citenamefont {Awaka}, \citenamefont {Kijima},
  \citenamefont {Takahashi}, \citenamefont {Hayakawa},\ and\ \citenamefont
  {Akimoto}}]{awaka-2009-synth-li6cal-li6bal}%
  \BibitemOpen
  \bibfield  {author} {\bibinfo {author} {\bibfnamefont {J.}~\bibnamefont
  {Awaka}}, \bibinfo {author} {\bibfnamefont {N.}~\bibnamefont {Kijima}},
  \bibinfo {author} {\bibfnamefont {Y.}~\bibnamefont {Takahashi}}, \bibinfo
  {author} {\bibfnamefont {H.}~\bibnamefont {Hayakawa}}, \ and\ \bibinfo
  {author} {\bibfnamefont {J.}~\bibnamefont {Akimoto}},\ }\href {\doibase
  10.1016/j.ssi.2008.10.022} {\bibfield  {journal} {\bibinfo  {journal} {Solid
  State Ionics}\ }\textbf {\bibinfo {volume} {180}},\ \bibinfo {pages} {602}
  (\bibinfo {year} {2009}{\natexlab{b}})}\BibitemShut {NoStop}%
\bibitem [{\citenamefont {Peng}\ \emph {et~al.}(2013)\citenamefont {Peng},
  \citenamefont {Wu},\ and\ \citenamefont {Xiao}}]{peng-2013-low-li5la3}%
  \BibitemOpen
  \bibfield  {author} {\bibinfo {author} {\bibfnamefont {H.}~\bibnamefont
  {Peng}}, \bibinfo {author} {\bibfnamefont {Q.}~\bibnamefont {Wu}}, \ and\
  \bibinfo {author} {\bibfnamefont {L.}~\bibnamefont {Xiao}},\ }\href {\doibase
  10.1007/s10971-013-2984-y} {\bibfield  {journal} {\bibinfo  {journal}
  {Journal of Sol-Gel Science and Technology}\ }\textbf {\bibinfo {volume}
  {66}},\ \bibinfo {pages} {175} (\bibinfo {year} {2013})}\BibitemShut
  {NoStop}%
\bibitem [{\citenamefont {O'Callaghan}\ \emph {et~al.}(2006)\citenamefont
  {O'Callaghan}, \citenamefont {Lynham}, \citenamefont {Cussen},\ and\
  \citenamefont {Chen}}]{ocallaghan-2006-struc-ionic}%
  \BibitemOpen
  \bibfield  {author} {\bibinfo {author} {\bibfnamefont {M.~P.}\ \bibnamefont
  {O'Callaghan}}, \bibinfo {author} {\bibfnamefont {D.~R.}\ \bibnamefont
  {Lynham}}, \bibinfo {author} {\bibfnamefont {E.~J.}\ \bibnamefont {Cussen}},
  \ and\ \bibinfo {author} {\bibfnamefont {G.~Z.}\ \bibnamefont {Chen}},\
  }\href {\doibase 10.1021/cm060992t} {\bibfield  {journal} {\bibinfo
  {journal} {Chemistry of Materials}\ }\textbf {\bibinfo {volume} {18}},\
  \bibinfo {pages} {4681} (\bibinfo {year} {2006})}\BibitemShut {NoStop}%
\bibitem [{\citenamefont {Novikov}\ \emph {et~al.}(2005)\citenamefont
  {Novikov}, \citenamefont {Kozinsky},\ and\ \citenamefont
  {Levitov}}]{novikov-2005-correl}%
  \BibitemOpen
  \bibfield  {author} {\bibinfo {author} {\bibfnamefont {D.}~\bibnamefont
  {Novikov}}, \bibinfo {author} {\bibfnamefont {B.}~\bibnamefont {Kozinsky}}, \
  and\ \bibinfo {author} {\bibfnamefont {L.}~\bibnamefont {Levitov}},\
  }\href@noop {} {\bibfield  {journal} {\bibinfo  {journal} {Physical Review
  B}\ }\textbf {\bibinfo {volume} {72}},\ \bibinfo {pages} {235331} (\bibinfo
  {year} {2005})}\BibitemShut {NoStop}%
\bibitem [{\citenamefont {Nozaki}\ \emph {et~al.}(2014)\citenamefont {Nozaki},
  \citenamefont {Harada}, \citenamefont {Ohta}, \citenamefont {Watanabe},
  \citenamefont {Miyake}, \citenamefont {Ikedo}, \citenamefont {Jalarvo},
  \citenamefont {Mamontov},\ and\ \citenamefont
  {Sugiyama}}]{nozaki-2014-li-qens}%
  \BibitemOpen
  \bibfield  {author} {\bibinfo {author} {\bibfnamefont {H.}~\bibnamefont
  {Nozaki}}, \bibinfo {author} {\bibfnamefont {M.}~\bibnamefont {Harada}},
  \bibinfo {author} {\bibfnamefont {S.}~\bibnamefont {Ohta}}, \bibinfo {author}
  {\bibfnamefont {I.}~\bibnamefont {Watanabe}}, \bibinfo {author}
  {\bibfnamefont {Y.}~\bibnamefont {Miyake}}, \bibinfo {author} {\bibfnamefont
  {Y.}~\bibnamefont {Ikedo}}, \bibinfo {author} {\bibfnamefont {N.~H.}\
  \bibnamefont {Jalarvo}}, \bibinfo {author} {\bibfnamefont {E.}~\bibnamefont
  {Mamontov}}, \ and\ \bibinfo {author} {\bibfnamefont {J.}~\bibnamefont
  {Sugiyama}},\ }\href {\doibase 10.1016/j.ssi.2013.10.014} {\bibfield
  {journal} {\bibinfo  {journal} {Solid State Ionics}\ }\textbf {\bibinfo
  {volume} {262}},\ \bibinfo {pages} {585} (\bibinfo {year}
  {2014})}\BibitemShut {NoStop}%
\bibitem [{\citenamefont {Awaka}\ \emph {et~al.}(2011)\citenamefont {Awaka},
  \citenamefont {Takashima}, \citenamefont {Kataoka}, \citenamefont {Kijima},
  \citenamefont {Idemoto},\ and\ \citenamefont
  {Akimoto}}]{awaka-2011-cryst-struc}%
  \BibitemOpen
  \bibfield  {author} {\bibinfo {author} {\bibfnamefont {J.}~\bibnamefont
  {Awaka}}, \bibinfo {author} {\bibfnamefont {A.}~\bibnamefont {Takashima}},
  \bibinfo {author} {\bibfnamefont {K.}~\bibnamefont {Kataoka}}, \bibinfo
  {author} {\bibfnamefont {N.}~\bibnamefont {Kijima}}, \bibinfo {author}
  {\bibfnamefont {Y.}~\bibnamefont {Idemoto}}, \ and\ \bibinfo {author}
  {\bibfnamefont {J.}~\bibnamefont {Akimoto}},\ }\href@noop {} {\bibfield
  {journal} {\bibinfo  {journal} {Chemistry Letters}\ }\textbf {\bibinfo
  {volume} {40}},\ \bibinfo {pages} {60} (\bibinfo {year} {2011})}\BibitemShut
  {NoStop}%
\end{thebibliography}%


\begin{thebibliography}{9}%
\makeatletter
\providecommand \@ifxundefined [1]{%
 \@ifx{#1\undefined}
}%
\providecommand \@ifnum [1]{%
 \ifnum #1\expandafter \@firstoftwo
 \else \expandafter \@secondoftwo
 \fi
}%
\providecommand \@ifx [1]{%
 \ifx #1\expandafter \@firstoftwo
 \else \expandafter \@secondoftwo
 \fi
}%
\providecommand \natexlab [1]{#1}%
\providecommand \enquote  [1]{``#1''}%
\providecommand \bibnamefont  [1]{#1}%
\providecommand \bibfnamefont [1]{#1}%
\providecommand \citenamefont [1]{#1}%
\providecommand \href@noop [0]{\@secondoftwo}%
\providecommand \href [0]{\begingroup \@sanitize@url \@href}%
\providecommand \@href[1]{\@@startlink{#1}\@@href}%
\providecommand \@@href[1]{\endgroup#1\@@endlink}%
\providecommand \@sanitize@url [0]{\catcode `\\12\catcode `\$12\catcode
  `\&12\catcode `\#12\catcode `\^12\catcode `\_12\catcode `\%12\relax}%
\providecommand \@@startlink[1]{}%
\providecommand \@@endlink[0]{}%
\providecommand \url  [0]{\begingroup\@sanitize@url \@url }%
\providecommand \@url [1]{\endgroup\@href {#1}{\urlprefix }}%
\providecommand \urlprefix  [0]{URL }%
\providecommand \Eprint [0]{\href }%
\providecommand \doibase [0]{http://dx.doi.org/}%
\providecommand \selectlanguage [0]{\@gobble}%
\providecommand \bibinfo  [0]{\@secondoftwo}%
\providecommand \bibfield  [0]{\@secondoftwo}%
\providecommand \translation [1]{[#1]}%
\providecommand \BibitemOpen [0]{}%
\providecommand \bibitemStop [0]{}%
\providecommand \bibitemNoStop [0]{.\EOS\space}%
\providecommand \EOS [0]{\spacefactor3000\relax}%
\providecommand \BibitemShut  [1]{\csname bibitem#1\endcsname}%
\let\auto@bib@innerbib\@empty
\bibitem [{\citenamefont {Giannozzi}\ \emph {et~al.}(2009)\citenamefont
  {Giannozzi}, \citenamefont {Baroni}, \citenamefont {Bonini}, \citenamefont
  {Calandra}, \citenamefont {Car}, \citenamefont {Cavazzoni}, \citenamefont
  {Ceresoli}, \citenamefont {Chiarotti}, \citenamefont {Cococcioni},
  \citenamefont {Dabo}, \citenamefont {Corso}, \citenamefont {Gironcoli},
  \citenamefont {Fabris}, \citenamefont {Fratesi}, \citenamefont {Gebauer},
  \citenamefont {Gerstmann}, \citenamefont {Gougoussis}, \citenamefont
  {Kokalj}, \citenamefont {Lazzeri}, \citenamefont {Martin-Samos},
  \citenamefont {Marzari}, \citenamefont {Mauri}, \citenamefont {Mazzarello},
  \citenamefont {Paolini}, \citenamefont {Pasquarello}, \citenamefont
  {Paulatto}, \citenamefont {Sbraccia}, \citenamefont {Scandolo}, \citenamefont
  {Sclauzero}, \citenamefont {Seitsonen}, \citenamefont {Smogunov},
  \citenamefont {Umari},\ and\ \citenamefont
  {Wentzcovitch}}]{giannozzi-2009-quant-espres}%
  \BibitemOpen
  \bibfield  {author} {\bibinfo {author} {\bibfnamefont {P.}~\bibnamefont
  {Giannozzi}}, \bibinfo {author} {\bibfnamefont {S.}~\bibnamefont {Baroni}},
  \bibinfo {author} {\bibfnamefont {N.}~\bibnamefont {Bonini}}, \bibinfo
  {author} {\bibfnamefont {M.}~\bibnamefont {Calandra}}, \bibinfo {author}
  {\bibfnamefont {R.}~\bibnamefont {Car}}, \bibinfo {author} {\bibfnamefont
  {C.}~\bibnamefont {Cavazzoni}}, \bibinfo {author} {\bibfnamefont
  {D.}~\bibnamefont {Ceresoli}}, \bibinfo {author} {\bibfnamefont
  {L.}~\bibnamefont {Chiarotti}, \bibfnamefont {Guido}}, \bibinfo {author}
  {\bibfnamefont {M.}~\bibnamefont {Cococcioni}}, \bibinfo {author}
  {\bibfnamefont {I.}~\bibnamefont {Dabo}}, \bibinfo {author} {\bibfnamefont
  {A.~D.}\ \bibnamefont {Corso}}, \bibinfo {author} {\bibfnamefont {S.~d.}\
  \bibnamefont {Gironcoli}}, \bibinfo {author} {\bibfnamefont {S.}~\bibnamefont
  {Fabris}}, \bibinfo {author} {\bibfnamefont {G.}~\bibnamefont {Fratesi}},
  \bibinfo {author} {\bibfnamefont {R.}~\bibnamefont {Gebauer}}, \bibinfo
  {author} {\bibfnamefont {U.}~\bibnamefont {Gerstmann}}, \bibinfo {author}
  {\bibfnamefont {C.}~\bibnamefont {Gougoussis}}, \bibinfo {author}
  {\bibfnamefont {A.}~\bibnamefont {Kokalj}}, \bibinfo {author} {\bibfnamefont
  {M.}~\bibnamefont {Lazzeri}}, \bibinfo {author} {\bibfnamefont
  {L.}~\bibnamefont {Martin-Samos}}, \bibinfo {author} {\bibfnamefont
  {N.}~\bibnamefont {Marzari}}, \bibinfo {author} {\bibfnamefont
  {F.}~\bibnamefont {Mauri}}, \bibinfo {author} {\bibfnamefont
  {R.}~\bibnamefont {Mazzarello}}, \bibinfo {author} {\bibfnamefont
  {S.}~\bibnamefont {Paolini}}, \bibinfo {author} {\bibfnamefont
  {A.}~\bibnamefont {Pasquarello}}, \bibinfo {author} {\bibfnamefont
  {L.}~\bibnamefont {Paulatto}}, \bibinfo {author} {\bibfnamefont
  {C.}~\bibnamefont {Sbraccia}}, \bibinfo {author} {\bibfnamefont
  {S.}~\bibnamefont {Scandolo}}, \bibinfo {author} {\bibfnamefont
  {G.}~\bibnamefont {Sclauzero}}, \bibinfo {author} {\bibfnamefont {A.~P.}\
  \bibnamefont {Seitsonen}}, \bibinfo {author} {\bibfnamefont {A.}~\bibnamefont
  {Smogunov}}, \bibinfo {author} {\bibfnamefont {P.}~\bibnamefont {Umari}}, \
  and\ \bibinfo {author} {\bibfnamefont {R.~M.}\ \bibnamefont {Wentzcovitch}},\
  }\href {http://stacks.iop.org/0953-8984/21/i=39/a=395502} {\bibfield
  {journal} {\bibinfo  {journal} {Journal of Physics: Condensed Matter}\
  }\textbf {\bibinfo {volume} {21}},\ \bibinfo {pages} {395502} (\bibinfo
  {year} {2009})}\BibitemShut {NoStop}%
\bibitem [{\citenamefont {Vanderbilt}(1990)}]{vanderbilt-1990-soft}%
  \BibitemOpen
  \bibfield  {author} {\bibinfo {author} {\bibfnamefont {D.}~\bibnamefont
  {Vanderbilt}},\ }\href {http://link.aps.org/doi/10.1103/PhysRevB.41.7892}
  {\bibfield  {journal} {\bibinfo  {journal} {Physical Review B}\ }\textbf
  {\bibinfo {volume} {41}},\ \bibinfo {pages} {7892} (\bibinfo {year}
  {1990})}\BibitemShut {NoStop}%
\bibitem [{\citenamefont {Perdew}\ \emph {et~al.}(1996)\citenamefont {Perdew},
  \citenamefont {Ernzerhof},\ and\ \citenamefont {Burke}}]{perdew-1996-ration}%
  \BibitemOpen
  \bibfield  {author} {\bibinfo {author} {\bibfnamefont {J.~P.}\ \bibnamefont
  {Perdew}}, \bibinfo {author} {\bibfnamefont {M.}~\bibnamefont {Ernzerhof}}, \
  and\ \bibinfo {author} {\bibfnamefont {K.}~\bibnamefont {Burke}},\ }\href
  {http://dx.doi.org/10.1063/1.472933} {\bibfield  {journal} {\bibinfo
  {journal} {The Journal of Chemical Physics}\ }\textbf {\bibinfo {volume}
  {105}},\ \bibinfo {pages} {9982} (\bibinfo {year} {1996})}\BibitemShut
  {NoStop}%
\bibitem [{\citenamefont {Broyden}(1970)}]{broyden-1970-conver-class}%
  \BibitemOpen
  \bibfield  {author} {\bibinfo {author} {\bibfnamefont {C.~G.}\ \bibnamefont
  {Broyden}},\ }\href {\doibase 10.1093/imamat/6.1.76} {\bibfield  {journal}
  {\bibinfo  {journal} {IMA Journal of Applied Mathematics}\ }\textbf {\bibinfo
  {volume} {6}},\ \bibinfo {pages} {76} (\bibinfo {year} {1970})}\BibitemShut
  {NoStop}%
\bibitem [{\citenamefont {Fletcher}(1970)}]{fletcher-1970}%
  \BibitemOpen
  \bibfield  {author} {\bibinfo {author} {\bibfnamefont {R.}~\bibnamefont
  {Fletcher}},\ }\href {\doibase 10.1093/comjnl/13.3.317} {\bibfield  {journal}
  {\bibinfo  {journal} {The Computer Journal}\ }\textbf {\bibinfo {volume}
  {13}},\ \bibinfo {pages} {317} (\bibinfo {year} {1970})}\BibitemShut
  {NoStop}%
\bibitem [{\citenamefont {Verlet}(1967)}]{verlet-1967-comput-exper}%
  \BibitemOpen
  \bibfield  {author} {\bibinfo {author} {\bibfnamefont {L.}~\bibnamefont
  {Verlet}},\ }\href {http://link.aps.org/doi/10.1103/PhysRev.159.98}
  {\bibfield  {journal} {\bibinfo  {journal} {Physical Review}\ }\textbf
  {\bibinfo {volume} {159}},\ \bibinfo {pages} {98} (\bibinfo {year}
  {1967})}\BibitemShut {NoStop}%
\bibitem [{\citenamefont {Hoover}(1985)}]{hoover-1985-canon}%
  \BibitemOpen
  \bibfield  {author} {\bibinfo {author} {\bibfnamefont {W.~G.}\ \bibnamefont
  {Hoover}},\ }\href {http://link.aps.org/doi/10.1103/PhysRevA.31.1695}
  {\bibfield  {journal} {\bibinfo  {journal} {Physical Review A}\ }\textbf
  {\bibinfo {volume} {31}},\ \bibinfo {pages} {1695} (\bibinfo {year}
  {1985})}\BibitemShut {NoStop}%
\bibitem [{\citenamefont {Bush}\ \emph {et~al.}(1994)\citenamefont {Bush},
  \citenamefont {Gale}, \citenamefont {Catlow},\ and\ \citenamefont
  {Battle}}]{bush-1994-self}%
  \BibitemOpen
  \bibfield  {author} {\bibinfo {author} {\bibfnamefont {T.~S.}\ \bibnamefont
  {Bush}}, \bibinfo {author} {\bibfnamefont {J.~D.}\ \bibnamefont {Gale}},
  \bibinfo {author} {\bibfnamefont {C.~R.~A.}\ \bibnamefont {Catlow}}, \ and\
  \bibinfo {author} {\bibfnamefont {P.~D.}\ \bibnamefont {Battle}},\ }\href
  {http://dx.doi.org/10.1039/JM9940400831} {\bibfield  {journal} {\bibinfo
  {journal} {Journal of Materials Chemistry}\ }\textbf {\bibinfo {volume}
  {4}},\ \bibinfo {pages} {831} (\bibinfo {year} {1994})}\BibitemShut {NoStop}%
\bibitem [{\citenamefont {M.~Woodley}\ \emph {et~al.}(1999)\citenamefont
  {M.~Woodley}, \citenamefont {D.~Battle}, \citenamefont {D.~Gale},\ and\
  \citenamefont {Richard A.~Catlow}}]{woodley-1999}%
  \BibitemOpen
  \bibfield  {author} {\bibinfo {author} {\bibfnamefont {S.}~\bibnamefont
  {M.~Woodley}}, \bibinfo {author} {\bibfnamefont {P.}~\bibnamefont
  {D.~Battle}}, \bibinfo {author} {\bibfnamefont {J.}~\bibnamefont {D.~Gale}},
  \ and\ \bibinfo {author} {\bibfnamefont {C.}~\bibnamefont {Richard
  A.~Catlow}},\ }\href {http://dx.doi.org/10.1039/A901227C} {\bibfield
  {journal} {\bibinfo  {journal} {Physical Chemistry Chemical Physics}\
  }\textbf {\bibinfo {volume} {1}},\ \bibinfo {pages} {2535} (\bibinfo {year}
  {1999})}\BibitemShut {NoStop}%
\end{thebibliography}%
\end{document}


\section{Supporting Information: Effects of Sublattice Symmetry and Frustration on Ionic Transport in Garnet Solid Electrolytes}


\section{Computational Methodology}
\label{sec-1}

Ab-initio simulations were performed in the framework of density functional theory (DFT) as implemented in the Quantum Espresso software \cite{giannozzi-2009-quant-espres}. Ultrasoft pseudopotentials \cite{vanderbilt-1990-soft} were used to represent ions, and the electronic wave functions were expanded on a plane-wave basis set with cut-off energy of 30 Ry. The Brillouin zone was sampled using only the $\Gamma$-point, which was sufficiently accurate for unit cells of about 100 atoms. The Kohn-Sham equations were solved self-consistently, employing a generalized gradient approximation \cite{perdew-1996-ration}, with a convergence criterion of 10$^{\text{-7}}$ Ry for the total energy. Structural relaxations were performed using the BFGS algorithm \cite{broyden-1970-conver-class,fletcher-1970} with the convergence threshold for forces of 10$^{\text{-3}}$ Ry/bohr. In ab-initio molecular dynamics (AIMD), the equations of motion were integrated using the leapfrog Verlet algorithm \cite{verlet-1967-comput-exper}. A time-step of 3 fs was used, and a Nose-Hoover thermostat \cite{hoover-1985-canon} was applied to regulate the temperature range (500 - 1300 K) and to reproduce the thermodynamics of a canonical (NVT) ensemble. Interatomic interactions were modeled with rigid-ion potentials which consist of the Coulomb interaction and short-range Buckingham potentials for Li-O and O-O interactions \cite{bush-1994-self} and La-O and Zr-O interactions \cite{woodley-1999}. In order to obtain good thermodynamics and hopping statistics the classical molecular dynamics (CMD) simulations were performed in a 2\texttimes{}2\texttimes{}2 cubic supercell for a time period of 10 ns with a time-step of 1 fs. A pre-equilibration simulation of 2 ns was conducted prior to the 10 ns production run. At 300 - 400 K, the diffusion coefficients (D) extracted from these simulations are small (D \textless{} 10-11 cm$^{\text{2}}$/s) and are associated with large relative errors.

In all CMD simulations the cell geometry was taken to be cubic and the cation composition was fixed; only the concentration of Li-ions was varied in order to study its influence on diffusion. Due to the approximate potentials and non-stoichiometric compositions used, the absolute values are not expected to describe the actual diffusion coefficients quantitatively; only qualitative comparison between different compositions is expected to be relevant. In contrast, in AIMD, the cation types were explicitly adjusted to charge-balance the cell.

Equation \eqref{diffusion} shows the Einstein relation that allows extraction of the diffusion coefficient ($D$) from the mean-squared displacement of the \ce{Li^+} ions. The mean square displacement (MSD) tracks the extent of diffusion of \ce{Li^+} within the crystal by calculating the deviation of the positions of the \ce{Li^+} at each time-step relative to a reference position.

\begin{equation}
D = \lim_{x \to \infty} \frac{1}{6t}  \langle | r(t) - r(0) |^{2} \rangle \label{diffusion}
\end{equation}

Based on the Einstein relation, the diffusion coefficient is obtained from a linear fit of the MSD. The slope-extracted diffusion coefficient is related to the activation energy based on the empirical Arrhenius equation \eqref{arrhenius}.

\begin{equation}
D = D_{0} e^{(\frac{-E_{A}}{k_{B}T})} \label{arrhenius}
\end{equation}

The ionic conductivity ($\sigma$) was calculated from the diffusion coefficient (D) using the Nernst-Einstein equation as,

\begin{equation}
\sigma = nq^{2}\frac{D}{k_{B}T}
\end{equation}

Here $k_{B}$ is the Boltzmann constant, $T$ the temperature, $n$ the concentration of \ce{Li^+} ions whose charge is $q=1$.

\section{Symmetry Classification of Site Occupancies}
\label{sec-2}

\begin{figure}[H]
\centering
\includegraphics[width=4in]{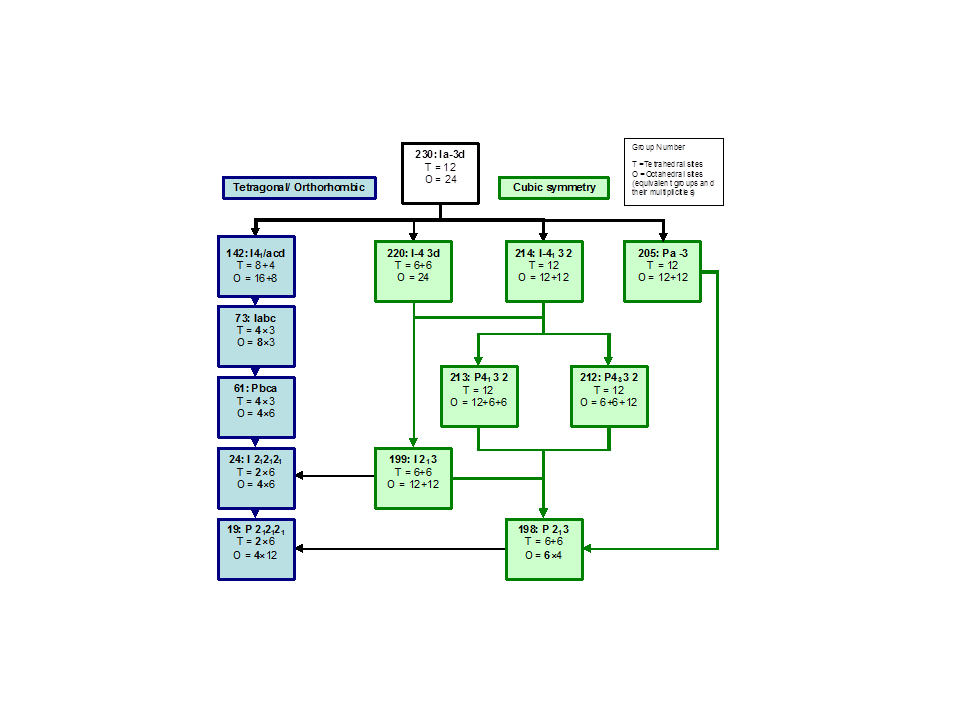}
\caption{Group-subgroup relationship tree that exemplifies a typical set of paths from the parent group 230 to lower symmetry subgroups. T and O represent tetrahedral and octahedral groups of sites, respectively, together with the multiplicities of each group. This set of relationships is not complete and only serves as an illustration of the symmetry projection procedure used to enumerate all possible ordered states. \label{fig:S1}}
\end{figure}

Fig. \ref{fig:S1} depicts the generalized symmetry classification of the garnet crystal. In the ideal cubic structure (group 230) \ce{Li^+} can potentially occupy three types of sites: octahedral (6-fold coordinated with O), tetrahedral (4-fold coordinated), and prismatic (6-fold coordinated). According to our DFT calculations, the structures where prismatic sites are occupied by Li are too high in energy to be relevant, so we do not consider them in our analysis. For the subsequent crystallographic subgroup projections, a few imperative symmetry considerations were made:

\begin{enumerate}
\item Starting with the highest symmetry, a high-symmetry \ce{Li^+} ordering is selected corresponding to the desired \ce{Li^+} content. For instance, in the case of \ce{Li7La3Zr2O12}, the 244-atom unit cell has 56 of the possible 72 octahedral and tetrahedral sites occupied. We then look for maximal subgroups of index 2 that would split the $48g$ and $24d$ Wyckoff positions into inequivalent groups of smaller multiplicities in such a way as to satisfy the occupancy ratio constraint (6 oct : 1 tet in the case of x=7). Symmetry sub-group analysis for $x=7$ reveals that it is impossible to find a cubic subgroup for the desired \ce{Li^+} occupancy but rather the highest symmetry attainable corresponded to the tetragonal \ce{Li7La3Zr2O12} (t-LLZ) structure (group No. 142) and several smaller tetragonal subgroups. This is also the case for $x=5$ and $x=4$.

\item For each \ce{Li^+} configuration corresponding to a particular subgroup and conjugacy class, a full atomic and crystal cell relaxation was performed using DFT. An energy comparison of the configurations allowed detection of the ground-state configuration (lowest-energy configuration) that is likely to be observed in experiment. We find that the lowest energy in all the ordered configurations identified in this way appear in the highest subgroup of parent prototype group No. 230 that satisfies the occupancy ration constraint. The excitation energies are estimated as the energy difference between the ground state and the next lowest-energy ordered configuration.
\end{enumerate}

\section{Ground state structures}
\label{sec-3}
In Tables \ref{tab:tab1} \ref{tab:tab2} and \ref{tab:tab3} we list the crystallographic representations of the ground state structures for $x=4$ and $x=6$ identified through the above procedure, combining symmetry projections and DFT optimization.

\begin{table}[H]
\caption{$x=9$ fully filled cubic garnet prototype structure (stoichiometry not balanced). Group No. 230. 
Unit Cell: 12.8 12.8 12.8 90.0 90.0 90.0. \label{tab:tab1}}
\centering
\begin{tabular}{llrrr}
Li & $24d$ & 0.250000 & 0.875000 & 0.000000\\
Li & $48g$ & 0.125000 & 0.682600 & 0.567400\\
A (La) & $24c$ & 0.125000 & 0.000000 & 0.250000\\
B (Zr) & $16a$ & 0.000000 & 0.000000 & 0.000000\\
O & $96h$ & 0.279650 & 0.105640 & 0.198940\\
\end{tabular}
\end{table}

\begin{table}[H]
\caption{$x=6$ cubic ordered state (\ce{Li6La3ZrTaO12}). Group No. 198. 
Unit Cell: 12.96  12.96  12.96  90.0 90.0 90.0 \label{tab:tab2}}
\centering
\begin{tabular}{llrrr}
Li(1)t & $12b$ & 0.62809565 & 0.51777692 & 0.24341914\\
Li(2)o & $12b$ & 0.42134654 & 0.3895379 & 0.6889208\\
Li(3)o & $12b$ & 0.86249175 & 0.33228876 & 0.43854475\\
Li(4)o & $12b$ & 0.55829781 & 0.69435057 & 0.34323682\\
La(1) & $12b$ & 0.12753211 & 0.0059228 & 0.25388513\\
La(2) & $12b$ & 0.88209927 & 0.99745617 & 0.74645235\\
Zr(1) & $4a$ & 0.0072975 & 0.0072975 & 0.0072975\\
Zr(2) & $4a$ & 0.7484067 & 0.2484067 & 0.2515933\\
Ta(1) & $4a$ & 0.24121324 & 0.74121324 & 0.75878676\\
Ta(2) & $4a$ & 0.50141341 & 0.50141341 & 0.50141341\\
O(1) & $12b$ & 0.30114506 & 0.12687324 & 0.19004257\\
O(2) & $12b$ & 0.85095734 & 0.52151886 & 0.04205643\\
O(3) & $12b$ & 0.72512746 & 0.91076796 & 0.80135235\\
O(4) & $12b$ & 0.13773853 & 0.46199925 & 0.94492937\\
O(5) & $12b$ & 0.78156338 & 0.59826182 & 0.71760309\\
O(6) & $12b$ & 0.34127734 & 0.02200982 & 0.57727936\\
O(7) & $12b$ & 0.22591701 & 0.39943203 & 0.31618047\\
O(8) & $12b$ & 0.64875228 & 0.97644017 & 0.48229974\\
\end{tabular}
\end{table}

\begin{table}[H]
\caption{$x=4$ tetragonal ground state. Group No. 95. 
Unit Cell: 12.77   12.77   12.64  90.0   90.0   90.0 \label{tab:tab3}}
\centering
\begin{tabular}{llrrr}
Li(1) & $4a$ & 0.000000 & 0.37862317 & 0.5000000\\
Li(2) & $4b$ & 0.500000 & 0.87135575 & 0.0000000\\
Li(3) & $8d$ & 0.59084348 & 0.67171652 & 0.44378762\\
Li(4) & $4a$ & 0.87671156 & 0.000000 & 0.7500000\\
Li(5) & $8d$ & 0.75902292 & 0.24510096 & 0.12652444\\
Li(6) & $4c$ & 0.42807484 & 0.42807484 & 0.1250000\\
La(1) & $4b$ & 0.88433456 & 0.500000 & 0.7500000\\
La(2) & $4a$ & 0.62821799 & 0.000000 & 0.7500000\\
La(3) & $4c$ & 0.75830768 & 0.75830768 & 0.6250000\\
La(4) & $4b$ & 0.62810149 & 0.500000 & 0.2500000\\
La(5) & $4a$ & 0.87522166 & 0.000000 & 0.2500000\\
La(6) & $4c$ & 0.74780442 & 0.25219558 & 0.375000\\
Ta & $8d$ & 0.750000 & 0.500000 & 0.500000\\
W & $8d$ & 0.750000 & 0.000000 & 0.000000\\
O(1) & $8d$ & 0.03199317 & 0.60993705 & 0.70125429\\
O(2) & $8d$ & 0.47879431 & 0.10228318 & 0.8045714\\
O(3) & $8d$ & 0.95223418 & 0.78006655 & 0.60957887\\
O(4) & $8d$ & 0.44813813 & 0.71893058 & 0.40179069\\
O(5) & $8d$ & 0.86011885 & 0.69897432 & 0.77979717\\
O(6) & $8d$ & 0.64868815 & 0.20280054 & 0.72515602\\
O(7) & $8d$ & 0.47251329 & 0.3968485 & 0.29770294\\
O(8) & $8d$ & 0.02826262 & 0.89200192 & 0.19808783\\
O(9) & $8d$ & 0.55585467 & 0.22400235 & 0.39348561\\
O(10) & $8d$ & 0.04964568 & 0.27821205 & 0.60520813\\
O(11) & $8d$ & 0.6421075 & 0.30037387 & 0.21039476\\
O(12) & $8d$ & 0.85755588 & 0.80240413 & 0.27346827\\
\end{tabular}
\end{table}

\bibliography{references}